\documentclass[12pt]{article}
\usepackage{amsfonts,amsmath,amssymb,graphics,epsfig}
\usepackage[greek,english]{babel}
\usepackage{graphicx}
\usepackage[iso-8859-7]{inputenc}

\begin{document}
\title{ {\large {\bf Weak Gravitational Field in Finsler - Randers Space and Raychaudhuri Equation }}   }
\author{{\large  P.C.Stavrinos }\footnote{pstavrin@math.uoa.gr},\\
{\small \emph{Department of Mathematics University of Athens 15784 Greece}}}
\date{}
\maketitle
\begin{abstract}
The linearized form of the metric of a Finsler - Randers space is studied in relation to the equations of motion, the deviation of geodesics and the generalized Raychaudhuri equation are given for a weak gravitational field. This equation is also derived in the framework of a tangent bundle.

By using Cartan or Berwald-like connections we get some types  \textquotedblleft gravito - electromagnetic\textquotedblright\thickspace curvature. In addition we investigate the conditions under which a definite Lagrangian in a Randers space leads to  Einstein field equations under the presence of electromagnetic field.

Finally, some applications of the weak field in a generalized Finsler spacetime for gravitational waves are given.
\end{abstract}

{\bf keywords:} Finsler geometry, weak gravitational field, Raychaudhuri equation.

%\tableofcontents
\newpage
\section{Introduction}
\qquad In the framework of general relativity, weak fields and gravitational waves have been studied by many authors decades ago e.g.\cite{J.Weber},\cite{R.Wagoner},\cite{K.Thorne and S. Kovacs}. One of the fundamental problems in general relativity is the study of gravitational waves. The existence of gravitational waves in  linear versions of the theory was already known in the early days of general relativity. First Einstein considered in a Minkowski spacetime with a metric $n_{\mu\nu}$ a small perturbation $\epsilon_{\mu\nu}$ such that the induced field $\alpha_{\mu\nu}=n_{\mu\nu}+\epsilon_{\mu\nu}$ with $\vert\epsilon_{\mu\nu}\vert\ll 1$ obeys by perturbation in linearized equations of motion. The linearized theory of gravity is an important theory because it can be utilized as a foundation for \textquotedblleft deriving\textquotedblright\thickspace General Relativity. By using the linearized field theory of gravitation some observable phenomena of our solar system and the universe can be detected
\cite{Farese},\cite{B.Sathyaprakash and B.Shutz} The weak field limit at a Finsler space-time has been studied in \cite{Calc},\cite{Stavrinos P.} and in the tangent bundle of a Finsler space by \cite{Pan.C.Stavrinos},\cite{V.Balan and P.C.Stavrinos},
\cite{N.Brinzei and S.Siparov}.

In our study we consider a pseudo- Finsler - Randers space-time of metric function \cite{G.Randers}
\begin{equation}
{\mathcal F}(x,y)=\sqrt{\alpha_{ij}(x)y^i y^j}+kA_i(x)y^i
\end{equation}
in the weak field limit where $\alpha_{ij}$ represents pseudo-Riemannian metric, $y^i=\frac{dx^i}{d\lambda}$(in applications $y^i$ represent velocity), $\lambda$ a parameter along the curve and $k$ a constant, $A_i$ represents the electromagnetic potential which is connected with the electromagnetic field by $F_{ij}=\frac{\partial A_j}{\partial x^i}-\frac{\partial A_i}{\partial x^j}$.

A Finsler-Randers space (FR) constitutes an important category of Finsler spaces from mathematical and physical perspective e.g.\cite{Bao D S-S Chern and Z Schen},\cite{S-K-S},\cite{Chang},\cite{Antonelli}.
In a FR space the condition of symmetry for the fundamental function ${\mathcal F}(x,y)$ is not satisfied ${\mathcal F}(x,y) \neq {\mathcal F}(x,-y)$. Causal considerations in pseudo-Finsler space-time that include only symmetries of fundamental functions \cite{Pfeifer C. and Wohlfarth M.} are very restricted and exclude by studying pseudo-FR spaces.\cite{Bao D S-S Chern and Z Schen},\cite{Antonelli},\cite{A.Bejancu and H.R. Farran}.

FR space can play a significant role in the theory of weak field and gravitational waves since ''a gravito - electromagnetic field'' is  intrinsically included in its metric. Einstein's General Relativity shows indeed that gravito - magnetic field may be associated with mass currents \cite{L.Iorio C.Corda},\cite{Martens}. As well a gravito - magnetic force was postulated as an explanation for the anomalous precession of Mercury's perihelion \cite{Mashhoon}. In addition, previous works \cite{Bel1}, \cite{Bel2}, \cite{Matte} ,\cite{Penrose}, \cite{Pirani} showed how the electric and magnetic parts of the curvature tensors were related to the electric and magnetic parts of the gravitational field as well as with gravitatational waves.\cite{C.Tsagas}\\

This paper is organized as follows: we deal with the linearized metric form of a pseudo FR spacetime in relation to the deviation of geodesics and Raychaudhuri equation. In this approach we extend a previous consideration which was given in \cite{Calc}.  By using Cartan and Berwald -like connections we get  \textquotedblleft gravito - electromagnetic curvatures\textquotedblright\thickspace for this space. We derive the equations of motion and the Raychaudhuri equation in the framework of a tangent bundle of a $n$-dimensional manifold $M$. We also give the linearized connection coefficients as well as establishing the Lorentz equation of the weak field. In addition we attribute some physical interpretations in the geometrical concepts under consideration.

Moreover in the framework of the weak gravitational field of a generalized Finsler spacetime some applications for gravitational waves are given.

\section{Linearized field theory of Randers space-time}\qquad The behavior of particles in a gravitational and electromagnetic field is expected to indicate that the physical geometry in the direction of a geometrical unification is the Finsler geometry.

In a Finsler space the metric function ${\mathcal F}(x,y)$ can be considered as a potential function since the metric tensor (gravitational potential)
\begin{equation}\label{first}
g_{ij}(x,y)=\frac{1}{2}\frac{\partial^2 {\mathcal F}^2}{\partial y^i\partial y^j}
\end{equation}
is produced by this function. The metric of a Randers space is given by virtue of (\ref{first}) in the form \cite{Bao D S-S Chern and Z Schen},\cite{Diakogiannis}.
\begin{equation}\label{randers}
g_{ij}=\alpha_{ij}+\frac{2k}{\sigma}y^s\alpha_{s(i}A_{j)}+k^2A_iA_j+\frac{k}{\sigma}y^lA_lm_{ij}
\end{equation}
where $\sigma=\sqrt{\alpha_{ij}y^iy^j},$ $m_{ij}=\alpha_{ij}-\sigma^{-2}\alpha_{is}\alpha_{jl}y^sy^l$ and $A_{(ij)}=\frac{1}{2}\left( A_{ij}+A_{ji}\right).$ We observe that the presence of an electromagnetic field in a region of spacetime breaks the isotropy and the description of spacetime is given by two metrics, one of which has a pseudo-Riemann structure $a_{ij}(x)$ that corresponds to a motion of a particle with mass m in the gravitational field and the second is metric of a charged particle of mass m that corresponds to a Finsler space of metric $g_{ij}(x,v)$ which represents a dynamical field. Connection coefficients of pseudo FR space are produced by those metrics. In a FR space the second term $y^i b_i$ can represent a measure of a cosmological anisotropy,a magnetic field or a spin-velocity. This consideration is analogous to Rosen's (1940) in which at each point of space-time a Euclidean and pseudo-Riemannian metric corresponds in each point of space-time.(Bimetric theory)

Finsler spaces are endowed with Cartan, Berwald connections and other different types of connections. Cartan connection has very important properties (metric compatibility) for models are closely related to standard physics \cite{Vacaru S. Stavrinos P. Gabourov E. Gontsa D.},\cite{Vacaru S.}.
Berwald connection is not generally compatible with the ,metric structure on total space, since it has "weak" compatibility only on the h-space on the tangent bundle. However for the case of a standard model extension a Berwald structure can be used for Lorentz violation in relation to gravitational waves e.g. \cite{A.Kostelecky}

The explicit form of Randers connection of Berwald-like coefficients  can be given in the form \cite{Asanov}
\begin{equation}\label{add}
L_{ij}^l=\alpha_{ij}^l+E_{ij}^l,
\end{equation}
where $\alpha_{ij}^l$ are the Riemannian Cristoffel symbols and $E_{ij}^l$ are given by
\begin{equation}\label{E}
E_{ij}^l=\frac{1}{2}\left(\alpha_{ij}y^kF_k^l+u_iF_j^l+u_jF^l_i\right)\alpha^{-1}-\frac{1}{2}u_iu_jy^kF^l_k\alpha^{-3}
\end{equation}
with $u_i=u_i(x,y)=\frac{\alpha_{ij}y^j}{\alpha(x,y)}\quad \alpha=\alpha(x,y)=\left(\alpha_{ij}(x)y^iy^j\right)^{1/2}.$ We note from (\ref{add}) and (\ref{E}) that the electromagnetic field enters in the connection coefficient of this space.

The geodesics of the Randers space are produced by the first variation of the action corresponding to the Lagrangian.
\begin{equation}
\frac{d}{d\lambda}\left(\frac{\partial{\mathcal F}}{\partial y^m}\right)-\frac{\partial{\mathcal F}}{\partial x^m}=0
\end{equation}
\begin{equation}\label{geo}
\frac{dy^m}{d\lambda}+L_{ij}^m(x,y)y^i y^j=0
\end{equation}
Because of (\ref{geo}) we get the well known Lorentz equation
\begin{equation} \label{lorentz}
\frac{d^2x^m}{d\lambda^2}+\alpha_{ij}^m\frac{d x^i}{d \lambda}\frac{d x^j}{d \lambda}+kF^m_j\frac{dx^j}{d\lambda}=0
\end{equation}
Eq.(\ref{lorentz}) represents the equation of motion of a charged particle in a gravitational and electromagnetic field, where $\lambda$ represents an affine parameter.

The equations of motion (\ref{lorentz}) under the perturbations $h_{ij}$ can be written in the form
\begin{equation}\label{lorentz2}
\frac{d^2x^m}{d\lambda^2}+\tilde{\alpha}_{ij}^m\frac{d x^i}{d \lambda}\frac{d x^j}{d \lambda}+kF^m_j\frac{dx^j}{d\lambda}=0
\end{equation}
where $\tilde{\alpha}_{ij}^m=\frac{1}{2}n^{ml}\left(\partial_i h_{jl}+\partial_j h_{im}-\partial_l h_{ij} \right)$ are the connection coefficients of the weak metric $\tilde\alpha_{ij}.$ The Eq. (\ref{lorentz2}) is useful for studying gravitational waves in a pseudo-FR space.

The curvature tensor of a Randers space can be produced by using Berwald-like connection coefficients in analogous to the form \cite{Asanov} for a weak field
\begin{equation}\label{curv}
H^i_{hjk}=R^i_{hjk}+E^i_{hjk}
\end{equation}
where $R^i_{hjk}$ is the Riemannian curvature tensor and $E^i_{hjk}$ is given by
\begin{equation}\label{ecurv}
\begin{split}
E^i_{hjk} & =\frac{1}{2}Q_{[jk]}\left( F^i_h F_{jk}+\alpha_{hk}F^m_j F^i_m-F_{hj}F^i_k\right)\\
& +Q_{[jk]}\left[(u_h\nabla_k F^i_j+y^m\alpha_{hj}\nabla_k F^i_m+u_j\nabla_k F^i_h)\alpha^{-1}-y^m u_h u_j\alpha^{-3}\nabla_k F^i_m\right]
\end{split}
\end{equation}
\\ with $\alpha=\alpha(x,y)=(\alpha_{ij}y^iy^j) ^{1/2}.$\\
Relation (\ref{curv}) is  rewritten
\begin{equation}
H^i_{hjk}(x,y)=\Lambda^i_{hjk}(x)+Q_{[jk]}\left(u_h \nabla_k F^i_j +\cdots -\nabla_k F^i_m\right)
\end{equation}
with $\Lambda^i_{hjk}=R^i_{hjk}+\frac{1}{2}Q_{[jk]}\left( F^i_h F_{jk}+\cdots -F_{hj}F^i_k\right)$ and $Q_{[jk]}=\frac{1}{2}\left(Q_{kj}-Q_{jk}\right)$
Applying the condtion
\begin{equation}\label{condition}
Q_{[jk]}\left(u_h\nabla_k F^i_j+\dot{x}^m\alpha_{hj}\nabla_k F^i_m+u_j\nabla_k F^i_h\right)\alpha^{-1}-\dot{x}^m u_h u_j\alpha^{-3}\nabla_k F^i_m=0
\end{equation}
in (\ref{curv}) we get the Lagrangian of the classical gravitational and electromagnetic fields
\begin{equation}
\Lambda=\Lambda_{hji}^i\alpha^{hj}=R+kF_{mn}F^{mn}, \quad k \thickspace\text{constant}.
\end{equation}
The variation of action's integral
\begin{equation}
\delta I=\delta\int{\Lambda}\sqrt{\vert \det \alpha_{ij} \vert}d^4x
\end{equation}

leads to the weak field equations
\begin{equation}
\left(R_{mn}-\frac{1}{2}\tilde\alpha_{mn}R\right)+k\left(F_{nr}F^r_m-\frac{1}{4} \tilde{\alpha}_{mn}F_{rs}F^{rs}\right)=0
\end{equation}
which are Einstein - Maxwell field equations of the \textquotedblleft gravito - electromagnetic\textquotedblright\thickspace of the Randers space for the vacuum under the condition (\ref{condition}). These equations have the same form as in the Riemannian ansatz in the presence of an electromagnetic field.

From a physical point of view the curvature $H^i_{hjk}$ can be considered as a\textit{\textquotedblleft gravito - electromagnetic curvature \textquotedblright\thickspace } of the space. We can say that it involves a gravito - electromagnetic \textquotedblleft current\textquotedblright\thickspace source. The metric of a weak gravitational field can be decomposed into the flat Minkowski metric plus a small perturbation
\begin{equation}
\tilde\alpha_{ij}=n_{ij}+h_{ij},\quad \vert h_{ij}\vert\ll 1.
\end{equation} Under a linearized approach of the gravitational field, the Randers metric function can be written in the form of a first approximation of the Riemannian metric $\alpha_{ij}$
\begin{equation}
{\mathcal F}(x,y)=\sqrt{\left(n_{ij}+h_{ij}(x,v)\right)v^i v^j}+k A_iv^i
\end{equation}
where $v^i=dx^i/d\tau$ is the 4-velocity of the particle, $n_{ij}=diag(1,-1,-1,-1)$ is the Minkowski metric, $\vert h_{ij}\vert\ll 1$ represents small perturbations to the flat spacetime metric and $k$ is a constant. The linearized form of the metric tensor, as introduced by (\ref{randers}), becomes
\begin{equation}\label{lin}
g_{ij}=n_{ij}+h_{ij}+\frac{2k}{\acute{\sigma}}v^s n_{s(i}A_{j)}+k^2A_iA_j+\frac{k}{\acute{\sigma}}v^lA_l\theta_{ij}
\end{equation}
where $\acute{\sigma}=\sqrt{n_{ij}v^iv^j},\thickspace\theta_{ij}=n_{ij}-\acute{\sigma}^{-2}n_{si}n_{jl}v^s v^l$ and $a_{(ij)}=\frac{1}{2}\left(a_{ij}+a_{ji}\right).$

Considering in (\ref{lin}) the case where $v^s=(1,0,0,0)$ we get the Finslerian potential $g_{00}$ of the Randers space for a test material point in the static case
\begin{equation}
g_{00}=1+h_{00}+k^2\phi^2+2k\phi
\end{equation}
with $\phi=A_0.$ In the case of $\phi=0$ we get $g_{00}=1+h_{00}.$ This relation is useful in order to derive the Riemannian or Newtonian limit from the equation of motion in a Randers space. In this case the equation of motion has the form
\begin{equation}
\dot{v}^l+L^l_{00}v^0 v^0=\dot{v}^l+L^l_{00}=0
\end{equation}
with
\begin{equation}
L^\mu_{00}=-\frac{1}{2}n^{\mu\lambda}\partial_\lambda h_{00}.
\end{equation}
The full interpretation of $L^l_{ij}$ is given by (\ref{add}).

The Finslerian potential $g_{00}$ takes the value 1 for the values of electromagnetic vector potential
\begin{equation}
\phi_{1,2}=\left(-1\pm\left(1-h_{00}\right)^{1/2}\right)(k)^{-1}.
\end{equation}
The Cristoffel symbols and the curvature tensor of the linearized Randers space will take the following form. By using (\ref{add}) and (\ref{curv}) we get
\begin{equation}\label{linadd}
\tilde{L}^i_{lj}=\tilde{a}^i_{lj}+h^i_{lj}
\end{equation}
\begin{equation}\label{lincurv}
\tilde{H}^i_{ljk}=\tilde{R}^i_{ljk}+h^i_{ljk}
\end{equation}
where
\begin{equation}\label{epsilon}
\tilde{R}^i_{ljk}=\frac{1}{2}n^{is}\left(\partial^2_{[kl}h_{sj]}-\partial^2_{[js}h_{lk]}\right)
\end{equation}
$\tilde{a}^i_{jk}$ are the linearized Riemannian Christoffel symbols and the curvature tensor. From (\ref{linadd}) and (\ref{lincurv}) the rest terms will be given in the form
\begin{equation}
h^m_{ij}=\frac{1}{2}\left(n_{ij}v^k F^m_k+u_i F^m_j+u_j F^m_i\right)n^{-1}-\frac{1}{2}u_i u_j v^kF^m_kn^{-3}
\end{equation}
\begin{equation}\label{h}
\begin{split}
h^i_{hjk} & =\frac{1}{2}Q_{[jk]}\left(F^i_h F_{jk}+n_{hk}F^m_j F^i_m-F_{hj} F^i_k\right)\\
& +Q_{[jk]}\left(u_h\partial_k F^i_j + v^m n_{hj}\partial_k F^i_m + u_j\partial_k F^i_h\right)n^{-1}-v^m u_h u_j n^{-3}\partial_k F^i_m
\end{split}
\end{equation}
\\

By using Cartan covariant differentiation in a Randers space we can express the third curvature tensor of Cartan $\bar{R}^i_{jkl}$ in the form \cite{Yasuda-Shimada}
\begin{equation}
\begin{split}
\bar{R}^i_{jhk} &=R^i_{jhk}+D^i_{jh\vert k}-D^i_{jk\vert h}+D^i_{hk}D^m_{jk}-D^i_{mk}D^m_{jh}\\
&+C^i_{jm}(R^m_{0hk}+D^m_{0h\vert k}-D^m_{0k\vert h}+D^m_{sh}D^s_{0k}-D^m_{sk}D^s_{0h})
\end{split}
\end{equation}
where $R^i_{jhk}$ is the Riemannian curvature, the symbol $\vert$ represents the Cartan covariant derivative and $D^i_{jk}$ is the difference tensor of the Finslerian gravitational field given by
\begin{equation}
\begin{split}
D^i_{jk}=&k\ell^i A_{(j,k)}+\frac{1}{2}k\left(\omega^i_j A_{0,k}+\omega^i_k A_{0,j} -\omega_{jk} A_{0,s} g^{is}\right)-C^i_{jk}\\
&+g^{is}k\left(A_{[s,j]}\ell_k+A_{[s,k]}\ell_j\right)+\frac{k}{\tau}\alpha^{mt}\big[g^{is}(A_{([t,s])}+A_{[t,0]}p_s\tau)C_{jkm}\\
&-C^i_{km}(A_{[t,j]}+A_{[t,0]}p_j\tau)-C^i_{jm}(A_{[t,k]}+A_{[t,0]}p_k\tau)\big]
\end{split}
\end{equation}
where $\omega_{ij}(x,y)=\tau\left(a_{ij} -k^2A_iA_j\right), \tau={\mathcal F}/a^{1/2}$ and $\ell^i=y^i/{\mathcal F}.$ In a linearised form the Cartan curvature tensor is expressed by
\begin{equation}
\tilde{K}^i_{jhk} = \tilde{R}^i_{jhk}+\tilde{D}^i_{jh\vert k}-\tilde{D}^i_{jk\vert h}+\tilde{C}^i_{jm}\left(\tilde{R}^m_{0hk}+\tilde{D}^m_{0h\vert k}-\tilde{D}^m_{0k\vert h}\right)
\end{equation}
where $\tilde{R}^i_{jhk}$ is given by (\ref{epsilon}) and $\tilde{D}^i_{jh}, \tilde{C}^i_{jm}$ represent the weak difference tensor and the weak Cartan connection coefficients, $\tilde{C}_{ijk}=\frac{1}{2}\frac{\partial\tilde{g}_{ij}}{\partial y_k} $.\\In (31) we have ignored terms $\tilde{D}^._{..} \tilde{D}^._{..}$ because of the condition $\vert h_{ij}\vert\ll 1$. The Ricci tensor of the weak \textit{"gravito - electromagnetic"} field is given by
\begin{equation}\label{ricci}
\tilde{K}_{ij}=\tilde{K}^s_{ijs}.
\end{equation}
For a perfect fluid moving in a Randers space with Cartan curvature $K^i_{jhk}$ Einstein's equations can be given in the form of weak field
\begin{equation}
\tilde{K}_{il}=k\left(T_{il}(x,V(x))-\frac{1}{2}T^\kappa_\kappa g_{il}\right)
\end{equation}
with $\tilde{K}_{il}=\tilde{R}_{il}+E_{il}$, where $\tilde{K}_{il}$ is the Ricci tensor, $\tilde{R}_{il}$ is the Riemannian one, $E_{il}$ is the contraction of $E^i_{jkl}$ by (\ref{curv}) and $T_{il}$ the energy - momentum tensor of FR space. Randers type space-time in cosmological considerations for a weak anisotropic field $u^a$ with $\ ||u^a|| \ll 1$ has been studied in \cite{S-K-S}.
\section{Weak Deviation of geodesics.Raychaudhuri equation}
\qquad The deviation of geodesics play an important role in General Relativity and Gravitation. In the Finslerian space-time it has been studied from mathematical and physical point of view \cite{GS.Asanov and P.C.Stavrinos},\cite{P.Stavrinos},\cite{S.Rutz},\cite{V.Balan and Stavrinos P.}. In a pseudo-Randers space the deviation of geodesics can be expressed by using \textit{"gravito-electromagnetic" curvature} (\ref{curv})
\begin{equation}\label{deva}
\frac{\delta^2 \xi^i}{\delta\lambda^2}+\bar{H}^i_{jhk}(x,v)\xi^j v^h v^k=0
\end{equation}
where $\frac{\delta\xi^i}{\delta\lambda}=\xi^i_{\vert h}v^h$ and $\xi^i_{\vert h}=\frac{\partial\xi^i}{\partial x^h}+\bar{R}^i_{hk}(x,v)\xi^k$, with $\lambda$ affine parameter. $\bar{R}^i_{hk}$ are the Cartan connection coefficients, $\xi^i$ represents the deviation vector  and $v^k$ the tangent vectors of a geodesic surface included in the Randers spacetime. We note from (\ref{deva}) that the deviation equation has two terms. The first term corresponds to the gravitational deviation, that will be observed if there is no electromagnetic field and it is associated with the $R^i_{hjk}$ part of the curvature tensor. The other term corresponds to a mixed geometrical and electromagnetic deviation and is associated with the $E^i_{hjk}$ part of the curvature tensor. The second term of deviation is connected to the force that two freely falling charged particles would exert to each other. Studying this case in a Riemannian spacetime has the consequence that the force does not necessarily result as a natural geometric effect as it does in a Finsler spacetime. If the $E^i_{hjk}$ vanishes then the deviation equation is reduced to the well known one of the Riemannian spacetime, namely
\begin{equation}
\frac{\delta^2 z^i}{\delta u^2}+R^i_{jkl}z^j v^k v^l=0
\end{equation}
\qquad Physically this means that we have two freely falling particles in the tidal field $R^i_{hjk}$ of spacetime. In the case where $R^i_{hjk}$ vanishes, we infer that the first term of the Randers metric corresponds to a Minkowski metric and the Finsler - Randers space becomes v - locally Minkowski \cite{R.Miron and M.Anastasiei}.
\begin{equation}
\mathcal {F}(x,v)=\sqrt{n_{\mu\nu} v^\mu v^\nu}+k A_i(x)v^i
\end{equation}
The only force that influences the two charged particles is due to the presence of the charged electromagnetic field. The deviation equation takes the form
\begin{equation}\label{devb}
\frac{\delta^2 z^i}{\delta u^2}+E^i_{jkl}z^j v^k v^l=0
\end{equation}
where $E^i_{jkl}$ is given by (\ref{ecurv}).
In this case the geometrical properties of the field are characterized by a homogeneous and anisotropic space. The metric fundamental tensor depends only on the velocities, which produce the anisotropic properties of the curved Finsler spacetime. Consequently there exists a frame of reference, where $\bar{R}^i_{hjk}$ vanishes. Under these circumstances the geodesic coordinates can be introduced for particles moving along these geodesics.

In a cosmological consideration the formula(36) can be given by $\mathcal {F}(x,V)=\sqrt{n_{\mu\nu} V^\mu V^\nu}+k W_i(x)V^i$ where $V=\tilde{H} d$ represents cosmological velocity depending on the cosmological Hubble parameter $\tilde{H}$ which is defined in the anisotropic Randers space-time with $\tilde{H} = \sqrt{H^2+Hz_{t}}$ cf. [13], $W_i$ represents an anisotropic field, $Z_t$ the variation of anisotropy and $d$ the distance. Such a consideration can be provided by a Finslerian osculating geometrical framework.If this field $W_i$ comes from by a curl the geodesics of this model are Riemannian. Gravitational waves in locally anisotropic spaces generate polarization patterns of the cosmic microwave background.

It is well known that the gravitational waves are connected to the deviation of geodesics. In order to study the weak field limit of a Randers space related to the deviation of the charged particles it is necessary to take into account relations (\ref{lincurv}) and (\ref{deva}). This is reasonable since in order to detect a gravitational wave at least two particles are needed.\\
Thus the deviation of geodesics of the weak Randers space is written in the form
\begin{equation}
\frac{d^2z^i}{d\tau^2}+\tilde{h}^i_{ljm}\frac{dz^j}{d\tau}\frac{dx^l}{d\tau}\frac{dx^m}{d\tau}=0
\end{equation}
\begin{equation}\label{devc}
\frac{d^2z^i}{d\tau^2}+\left(\epsilon^i_{ljm}+h^i_{ljm}+\right) \frac{dx^l}{d\tau}\frac{dz^j}{d\tau}\frac{dx^m}{d\tau}=0
\end{equation}
If we consider our test particles to be moving slowly then we can express the 4-velocities as a unit vector in the time direction. Hence we write
\begin{equation}\label{vec}
\frac{dx^i}{d\tau}=(1,0,0,0).
\end{equation}
In order to compute the Riemannian tensor in a first approximation we get from (\ref{devc})
\begin{equation}
\epsilon^i_{0j0}=\frac{1}{2}n^{ik}\left(\epsilon_{jk,00}-\epsilon_{0j,k0}-\epsilon_{0k,0j}+\epsilon_{00,kj}\right)
\end{equation}
In our case $\epsilon_{i0}=0$ and the Riemannian tensor takes the form
\begin{equation}\label{e}
\epsilon^i_{0j0}=\frac{1}
{2}\epsilon^i_{j,00}.
\end{equation}
The second term of (\ref{devc}) $h^i_{ljm}$ because of (\ref{vec}) and (\ref{h}) becomes
\begin{equation}\label{h2}
h^i_{0j0}=2F^i_0F^0_{j}+u_j\partial_0F^i_0.
\end{equation}
Furthermore (\ref{devc}) will take the form because of (\ref{e}) and (\ref{h2})
\begin{equation}\label{devd}
\frac{\partial^2z^i}{\partial \tau^2}+\left(\frac{1}{2}\frac{\partial^2\epsilon^i_j}{\partial \tau^2}+2F^i_0F^0_{j}+u_j\frac{\partial F^i_0}{\partial \tau}\right)\frac{\partial z^j}{\partial \tau}=0.
\end{equation}
The equation (\ref{devd}) coincides with the corresponding equation for a weak field limit of the Riemannian case, which is given in its full form by
\begin{equation}\label{deve}
\frac{d^2n^\mu}{ds^2}+R^\mu_{\nu\kappa\lambda}n^\kappa v^\nu v^\lambda=\Phi^\mu
\end{equation}
with
\begin{equation}
\Phi^\mu=k\left(\frac{dF^\mu_\kappa}{ds}v^\kappa+F^\mu_\nu F^\nu_\kappa v^\kappa\right),\quad k:\text{constant.}
\end{equation}
In general $\Phi^\mu$ represents a non-gravitational force, for instance a spring.
The difference between (\ref{deve}) and (\ref{deva}) is that in (\ref{deve}) the electromagnetic field has been added {\it ad hoc}. This means that the term $\Phi^\mu$ plays the role of the interaction external force between two nearby charged masses, moving in non-geodesical paths. In the equation (\ref{deva}) of the Randers space the electromagnetic field is incorporated in the geometry. In this approach the two charged masses move in geodesics of the Finsler space and their relative acceleration is determined by the curvature of the gravitational and electromagnetic fields, which is produced by the energy momentum tensor. Randers-type spaces best express a profound relation between physics and geometry.

An extension of the geodesic deviation equations constitutes the Raychadhuri equation. Raychadhuri equation is of important significance in Relativity theory and Cosmology because of its connection with singularities e.g. \cite{Raychaudhuri.A},\cite{S.Hawking-Ellis},\cite{A.Kouretsis and C.Tsagas}. In Finsler-Randers space-time this equation has been studied in a previous paper \cite{P.C.Stavrinos}. Its form in the weak Finslerian limit is given by
\begin{equation}\label{ray}
\frac{d\tilde{\theta}}{d\tau}=-\frac{1}{3}\tilde{\theta}^2-\tilde{\sigma}_{ik} \tilde{\sigma}^{ik}+\tilde{\omega}_{ik}\tilde{\omega}^{ik}-K_{i\ell}V^i V^\ell +\dot{V}^{i}_{;i}
\end{equation}
where $K_{il}$ represents the weak Cartan tensor,$\tilde{\theta}, \tilde{\omega_{ik}},\tilde{\sigma}_{ik}$ are the expansion,vorticity and the shear are defined by the
following forms:
\begin{subequations}\label{tso}
\begin{align}
    & \widetilde{\theta}
    =\Lambda_{ij}h^{ij}=V^{i}_{|i}-C^{i}_{im}\dot{V}^m\\
    &\tilde{\omega}_{ik}=\Lambda_{[ik]}+\dot{V}_i V_k -\dot{V}_k V_i \\
    &\tilde{\sigma}_{ik}=\Lambda_{(ik)}-\frac{1}{3}\tilde{\theta}
    h_{ik}-2C_{ikm}V^m -\dot{V}_iV_k-\dot{V}_kV_i
\end{align}\end{subequations}
where $\Lambda_{(ik)}=V_{i;k}$ is the covariant derivative of the oscullating Riemannian space $V^i$ a unit vector $V^i V_{i} = 1$ In the case of geodesics the last term $\dot{V}^{i}_{;i}$ vanishes in (\ref{ray}) and (\ref{tso}). The introduction of Cartan tensor in (48) assigns an anisotropic structure for the Raychaudhuri equation. The linearized Raychaudhuri equation in a Randers spacetime in a first approach is expressed without vorticity and shear by
\begin{equation}\label{ray2}
\frac{d\tilde{\theta}}{d\tau}=-\frac{1}{3}\tilde{\theta}^2- \tilde{K}_{i\ell}V^i V^\ell
\end{equation}
In the case that $\dot{\tilde{\theta}}=0, \tilde{\sigma_{ij}}=0,\tilde{\omega}_{ij}=constant$, from
$\eqref{ray}$ the tidal field $K_{il}V^iV^l$ is due to the vorticity $\tilde{\omega}$ which plays the role of vacuum energy (cosmological
constant). It is analogous to a centrifugal field of the Newtonian theory. It counterbalances the tidal field.

\paragraph{\emph{Remark}} The fundamental sense of photon surfaces and their geometry has been defined and developed in [41],[42] for a timelike surface in a spherical symmetric space with determined properties.

In a pseudo-Finsler space-time $M$ with spherically symmetric metric [35] in which $\tilde{\sigma}_{ij}=0$ and $\tilde{\omega}_{ij}=0$, we can analogously consider a Finslerian photon surface $S$, where $S$ represents a timelike surface of $M$. Here the Raychaudhuri equation takes the form
\begin{equation}
\frac{d\theta_{(2)}}{d\tau} = -\frac{1}{2} \theta^2_{(2)} - K_{il}^{(3)} X^i X^l,
\end{equation}
where $\theta_{(2)}$ denotes the expansion of a vector field $X$ in the surface $S$, $\tau$ the affine parameter and $K_{il}^{(3)}$ the Ricci tensor of Cartan curvature.

On a physical viewpoint the anisotropic Cartan tensor is introduced in the geometry of spacetime because of a primordial vector field in the Finsler- Randers spacetime. Such a case has been studied in \cite{S-K-S} where the linearized Raychaudhuri equation stands
\begin{equation}
\frac{1}{3}\tilde{\theta}^2\tilde{q}=4\pi G(\mu+3P)\frac{H}{\tilde{H}}+f(\alpha,\dot\alpha,\ddot\alpha,z_t)
\end{equation}
where $\tilde{q}, \mu, p, \alpha, z_t$ represent the deceleration parameter, the density of matter, the pressure, the scale factor and the variation of anisotropy, which is connected to Cartan connection component $z_t=C_{000,0}$.

The Raychaudhuri equations can also be derived in the framework of a tangent bundle TM of a $n$-dimensional manifold by using of d-connection and the Ricci-identities. In this case we consider the d-curvature $R^{i}_{jkl}$ and the Ricci identities for a tangent horizontal vector field $X = X^H= X^i \delta/\delta x^i$ along a congruence of geodesics on TM [37]. So we have the relation

\begin{equation}
X^i_{|kl} - X^i_{|lk} = R^i_{jkl}X^j - T^h_{kl}X^i_{|h} - R^a_{kl}X^i|_a,
\end{equation}
or
\begin{equation}
X^l X^i_{|kl} = X^i_{|lk} X^l +  R^i_{jkl}X^jX^l - T^h_{kl}X^i_{|h}X^l - R^a_{kl}X^i|_aX^l.
\end{equation}
Because of geodesics the relation $(X^lX^i_{|l})_{|k} =0$ is valid so we have
\begin{equation}
X^l X^i_{|kl} = -X^l X_{|k}X^i_{|l} + R^i_{jkl} X^j X^l - T^h_{kl} X^i_{|h} X^l -R^a_{kl} X^i|_a X^l.
\end{equation}

Taking the trace of the previous equation we have
\begin{equation}
X^l\tilde{h}^k_i X^i_{|kl} = - X^l_{|k} X^i_{|l} \tilde{h}^k_i + R^i_{jkl} X^j X^l \tilde{h}^k_i -T^h_{kl} X^i_{|h} X^l \tilde{h}^k_i - R^a_{kl} X^i|_a X^l \tilde{h}^k_i. \label{eq:55}
\end{equation}

We decompose the $h$-covariant derivative with kinematical terms
\begin{equation}\label{eq:149}
X^i_{|l} = \frac{1}{n-1} \tilde{\Theta}\tilde{h}^i_l + \tilde{\sigma}^i_l + \tilde{\omega}^i_l,
\end{equation}
where $\tilde{\Theta}$, $\tilde{\sigma}$, $\tilde{\omega}$ represent the expansion, shear and vorticity for the extended congruence of geodesics on TM, which are defined as
\begin{align}
\tilde{\Theta} &= X^i_{|l} \tilde{h}^l_i,\label{eq:150}\\
\tilde{\sigma}_{il} &= X_{i|l}+X_{l|i} - \frac{1}{3} \tilde{\Theta} \tilde{h}_{il},\\
\tilde{\omega}_{il} &= X_{i|l} - X_{l|i},
\end{align}
with
\begin{equation}
\tilde{h}_{il} = g_{il} - X_iX_l
\end{equation}
the projection operator, $h_{il}$ gives us $\tilde{h}_{il} X^i =0$ for a normalized $X_i$.
In the above mentioned relations, \eqref{eq:149}, \eqref{eq:150}, we used
\begin{equation}
\tilde{h}^i_l = g^{ik}\tilde{h}_{lk}.
\end{equation}

We finally get from \eqref{eq:55}
\begin{multline}\label{eq:Raychadhuri}
X^l \tilde{\Theta}_{|l} = \frac{d\tilde{\Theta}}{d\tau} = R_{kl}X^k X^l - T^h_{li} X^i_{|h} X^l - R^a_{li}X^i|_a X^l- X^l_{|k} X^k_{|l} = R_{il} X^i X^l\\
    - T^h_{li} \left(\frac{1}{n-1} \tilde{\Theta}\tilde{h}^i_h + \tilde{\sigma}^i_h + \tilde{\omega}^i_h \right)X^l - R^a_{li} \left( \frac{1}{n-1} \Theta \tilde{h}^i_a + \sigma^i_a + \omega^i_a \right)X^l - \frac{1}{n-1} \tilde{\Theta} - \sigma^l_k \sigma^k_l - \omega^l_k\omega^k_l.
\end{multline}
where we put
\begin{displaymath}
X^i|_a = \frac{1}{n-1} \tilde{\Theta}\tilde{h}^i_a + \tilde{\sigma}^i_a + \tilde{\omega}^i_a,
\end{displaymath}
with $\tilde{h}^i_a = h^b_a \delta^i_c \delta^c_b$, $\sigma^i_a = \sigma^b_a \delta^i_c \delta^c_b$, $\omega^i_a = \omega^b_a \delta^i_c \delta^c_b$ and $\delta^i_c$ represent the generalized Kronecker symbols connecting with h-bases and v-bases. The equation \eqref{eq:Raychadhuri} is the Raychadhuri equation for the horizontal space of the tangent bundle.

\begin{equation}
R^a_{li} = \frac{\partial N^a_l}{\partial x^i} - \frac{\partial N^a_i}{\partial x^l}
\end{equation}
represents the curvature of non-linear connection.

By using of d-curvature $S^a_{bcd}$, the Ricci identities are written
\begin{equation}
X^a \vert_{bc} - X^a \vert_{cb} = S^a_{dbc} X^d - S^d_{bc} X^a \vert_{d},
\end{equation}
where the vector field $X=X^a \frac{\partial}{\partial y^a}$ belongs to the vertical space $S^a_{bc} = C^a_{bc} - C^a_{cb}$ represent the torsion and $C^a_{bc}$ the d-connection coefficients of the vertical space. In analogy to the consideration of Finslerian fluids cf.[43] we can get the Raychaudhuri equations. For the definitions of the decomposition of vertical covariant derivative of vertical geodesics expansion, shear and rotation we use the relations
\begin{equation}
X^a \vert_{b} = \frac{1}{3} \Theta h^a_{b} + \sigma^a_{b} + \omega^a_{b}
\end{equation}
where the expansion $\Theta$ is given by
\begin{equation}
\Theta = X^a \vert_{b} h^b_{a} = X^a \vert_{a},
\end{equation}
$h_{ab}(x,y)$ represents the v-metric on TM which is connected with the h-metric
$g_{ij}(x,y)=\delta^a_{i} \delta^b_{j} h_{ab} (x,y) \sigma_{ab}$ and $\omega_{ab}$. We define the shear and rotation by
\begin{align}
\sigma_{ab} &=X_a \vert_{b} +X_b \vert_{a} - \frac{1}{3} \Theta h_{ab},\\
\omega_{ab} &=X_{a} \vert_{b} - X_{b} \vert_{a}.
\end{align}

Because of the above mentioned relations one obtains an expression for the Raychaudhuri equation in the vertical space in the forms
\begin{multline}
X^c \Theta \vert_{c} = \frac{d\Theta}{d\tau}= S_{dc}X^d X^c - S^d_{ca} X^a \vert_{d} X^c - X^c \vert _{b} X^b \vert _{c}\\
    =-\frac{1}{n-1} \Theta^2 - \sigma^a_{b} \sigma^b_{a} - \omega^a_{b} \omega^b_{a} + S_{dc} X^d X^c - S^b_{ca} X^c \left(\frac{1}{n-1} \Theta h^a_{b} + \sigma^a_{b} + \omega^a_{b} \right)
\end{multline}
\\
The Raychaudhuri equation can also be derived on the tangent bundle of a Finsler-Randers space-time as well as for its weak field limit by considering the analogous curvature in the rel.~\eqref{eq:Raychadhuri}.
\\
\\
\\
{\bf Applications:}\\
\\
{\bf 1.}\quad In the $(x^0,x^1,x^2,x^3)$ coordinates of an inertial frame generalized Finsler metrics that are very close to the flat metric can be written
\begin{equation}\label{weak}
g_{\mu\nu}(x,y)=\eta_{\mu\nu}+h_{\mu\nu}(x,y)
\end{equation}
where $y=\frac{dx}{dt}$ and $h_{\mu\nu}(x,y)$ are small anisotropic perturbations to the flat spacetime. These metric perturbations describe a gravitational wave. The line - element for a plane gravitational wave spacetime can be expressed in the form
\begin{equation}\label{li}
ds^2=dt^2-\left(1+f[(x_0,x_3),y]dx^2_1\right)-\left(1-f[(x_0,x_3),y]dx^2_2\right)+dx_3^2
\end{equation}
where the function $\delta_{\mu\nu}f(x,y)=h_{\mu\nu}(x,y),$ with
\begin{equation}
\left\vert f[(x^0-x^3),y]\right\vert\ll 1.
\end{equation}
If the wave has a definite frequency $\omega$, amplitude $\alpha$ and phase $\delta$ we can write
\begin{equation}
f[(x^0-x^3),y]=\alpha\sin [\omega(x^0-x^3)+\delta]\thickspace y.
\end{equation}
In the case of the weak field limit of a Finslerian or generalized Finslerian space-time the vacuum field equation holds analogous form to Newton's gravity and General Relativity.

We specify a gauge under a coordinate transformation $x^\mu\rightarrow \bar{x}^\mu=x^\mu+\frac{1}{2}Q^\mu_{\;\; \alpha \beta}x^\alpha x^\beta$ where $Q^\mu_{\;\; \alpha \beta}\sim \tilde{L}^\mu_{\;\;\alpha\beta}\sim \epsilon$, $\tilde{L}^\mu_{\;\;\alpha\beta}$ represent the linearized Christoffel symbols given by (25). We choose a Lorentz gauge $\eta^{\mu\nu} \tilde{L}^\rho_{\;\;\mu\nu}=0$.

In the weak field limit the connection coefficients $h^\rho_{\;\;\mu\nu}$ must be zero, so we have $\tilde{L}^\rho_{\;\;\mu\nu}=\tilde{a}^\rho_{\;\;\mu\nu}$ which leads to the equivalent Lorentz gauge
\begin{equation}
\partial_\mu h^{\mu}_{\;\lambda}(x,y)-\frac{1}{2}\partial_\lambda h=0.
\end{equation}
By using the linearized Einstein equations  for the vacuum (27) with $\tilde{R}_{\mu\nu}=0$ we get the equivalent form of the wave equation in the new system of coordinates
\begin{equation}
\Box \bar{h}_{\mu\nu}(x,y)=0.
\label{gWEQ}
\end{equation}
A solution of (\ref{gWEQ}) is a gravitational wave
\begin{equation}
\bar{h}_{\mu\nu}(x,y)=\exp (ik_\sigma(y)x^\sigma)C_{\mu\nu}
\label{GWslt}
\end{equation}
with $C_{\mu\nu}=C_{\nu\mu}$ a constant polarization tensor and the wave vector $k_\mu (y)$ is a function of $y$ because of anisotropic metric (71).

By inserting (\ref{GWslt}) to (\ref{gWEQ}) we derive
\begin{equation}
\eta^{\mu\nu}k_{\mu}(y)k_{\nu}(y)=k^2=0
\end{equation}
with $k_\mu=(\omega,k_1,k_2,k_3)$. The plane wave (\ref{GWslt}) is therefore a solution to the linearized equations if the wave vector is null.

A linearized form of the curvature $\tilde{H}^i_{\;\;ljk}$ (relation (26)) can be produced by the linearized coefficients $\tilde{L}^i_{\;\; jk}$. In order to get gravitational waves the harmonic gauge condition is chosen as
\begin{equation}
\frac{\partial \tilde{h}^\mu_{\;\; \nu}}{\partial y^\mu}=0
\end{equation}
and traceless $\tilde{h}^\mu_{\;\; \nu}=0$.

The traceless and transverse (TT) gauge condition for the equation of motion is given by
\begin{equation}
\Box h^{TT}_{\;\;\mu\nu}(x,y)=0.
\end{equation}
A particular useful set of solutions to this wave equation are the plane waves
\begin{equation}
 h^{TT}_{\;\;\mu\nu}(x,y)=C'_{\mu\nu}\exp (ik_a(y)x^a)
\end{equation}
with $C'_{\mu\nu}$ a constant symmetric tensor. The plane wave $ h^{TT}_{\;\;\mu\nu}$ is a solution to the linearized wave equation if the wave vector is null $k^a(y)k_a(y)=0$.

 The geodesic deviation equation with the curvature $\tilde{H}^i_{\;\; jkl}$, for two particles with $\upsilon^\mu=(1,0,0,0)$ 4-velocity and separation vector  $\zeta^{x^2}$ in the $x^2$-direction implies
\begin{eqnarray}
\frac{\partial ^2 \zeta ^{x^2}}{\partial t^2}=\frac{1}{2}\frac{\partial ^2}{\partial t^2} h^{TT}_{\;\; x^2x^2} \\
\frac{\partial ^2 \zeta ^{x^1}}{\partial t^2}=\frac{1}{2}\frac{\partial ^2}{\partial t^2} h^{TT}_{\;\; x^1x^2}
\end{eqnarray}
These equations are important to describe a polarization of the gravitational wave in $y=(x^1,x^2)$ direction in the framework of the anisotropic metric (71).
\\
\\
{\bf 2.} In the previous application the coordinates $x^i, \thickspace i=0,1,2,3$ are independent of time. Distances between test masses in Euclidean plane can be calculated from $X(t), Y(t)$ coordinates by using a Randers metric. We can put the coordinates in the form
\begin{subequations}
\begin{align}
X &=(1+\frac{1}{2}\alpha\sin\omega t)\thickspace x^1\\
Y &=(1-\frac{1}{2}\alpha\sin\omega t)\thickspace x^2\\
\end{align}
\end{subequations}
then
\begin{subequations}
\begin{align}
\dot{X}&=\frac{dX}{dt}=\frac{1}{2}\alpha\omega\cos\omega t\thickspace x^1\\
\dot{Y}&=\frac{dY}{dt}=-\frac{1}{2}\alpha\omega\cos\omega t\thickspace x^2
\end{align}
\end{subequations}
A plane gravitational wave of the form (\ref{li}) with $f[(x^0-x^3),y^3]=\alpha \sin[\omega (x^0-x^3)]\thickspace y^3,$ and $y^3=(0,0,1,0)$ propagates in the $y^3$ direction. Some test masses in the $x-y$ plane are as rest in a circle about a central test mass. After the gravitational wave passes in time t the circle is squeezed in the Y-direction and expanded in the X-direction, therefore the circle is transformed to an elliptic shape.

We define the indicartix curve $I_p$ to be a circle with center 0 and radius $\sqrt{1+f(x,y)}\thickspace OP$ where $P(x,y)$ is an arbitary point and $f(x,y)$ is a positive valued function. We can apply the Randers metric in order to calculate the distances between the masses \cite{Antonelli}. Therefore we have
\begin{equation}
{\mathcal L}(x,y)=\frac{\lambda+\sqrt{f(x,y)\rho^2+\lambda^2}}{f(x,y)}
\end{equation}
where
\begin{equation}
\rho^2=\frac{\dot{X}^2+\dot{Y}^2}{X^2+Y^2}\quad \lambda=\frac{X\dot{X}+Y\dot{Y}}{X^2+Y^2}
\end{equation}
\\
{\bf Conclusions}
\\
\\
We studied the behaviour of particles moving in a gravitational and electromagnetic field with the physical geometry of a Finsler--Randers (FR) space. Cartan and Berwald connections are applied for studying a linearized version of a weak field limit in F-R spaces.

In virtue of curvature tensors of the space of considerations some physical characterizations and interpretations in the sense of a \guillemotleft gravito--electromagnetic curvature \guillemotright are given. Such a concept could play a role in the bending of light geodesics and gravitational lensing in a region of locally anisotropic space-time.

In paragraph 3 the Raychaudhuri equations are extended and they were derived in the framework of a tangent bundle. This consideration can give an additional interest to a string theory.

Finally, some applications of Randers metric for gravitational waves are presented.
\\
\\
{\bf Acknowledgement}
\\
\\
The author is grateful to the University of Athens (Special Accounts for Research Grants) for the support to this work.

\end{document}